\documentclass[a4paper,preprint]{aastex}

\usepackage{natbib}

\begin{document}

\title{A Search for Additional Planets in the NASA {\it EPOXI} Observations of the Exoplanet System GJ 436}

\author{Sarah~Ballard\altaffilmark{1}, Jessie~L.~Christiansen\altaffilmark{1}, David~Charbonneau\altaffilmark{1}, Drake~Deming\altaffilmark{2}, Matthew~J.~Holman\altaffilmark{1}, Daniel~Fabrycky\altaffilmark{1,3}, Michael~F.~A'Hearn\altaffilmark{4}, Dennis~D.~Wellnitz\altaffilmark{4}, Richard~K.~Barry\altaffilmark{2}, Marc~J.~Kuchner\altaffilmark{2}, Timothy~A.~Livengood\altaffilmark{2}, Tilak~Hewagama\altaffilmark{2,4}, Jessica~M.~Sunshine\altaffilmark{4}, Don~L.~Hampton\altaffilmark{5}, Carey~M.~Lisse\altaffilmark{6}, Sara~Seager\altaffilmark{7}, and Joseph~F.~Veverka\altaffilmark{8}}

\altaffiltext{1}{Harvard-Smithsonian Center for Astrophysics, 60 Garden Street, Cambridge, MA 02138, USA; sballard@cfa.harvard.edu}
\altaffiltext{2}{Goddard Space Flight Center, Greenbelt, MD 20771, USA}
\altaffiltext{3}{Michelson Fellow}
\altaffiltext{4}{University of Maryland, College Park, MD 20742, USA}
\altaffiltext{5}{University of Alaska Fairbanks, Fairbanks AK 99775, USA}
\altaffiltext{6}{Johns Hopkins University Applied Physics Laboratory, Laurel, MD 20723, USA}
\altaffiltext{7}{Massachusetts Institute of Technology, Cambridge, MA 02159, USA}
\altaffiltext{8}{Cornell University, Space Sciences Department, Ithaca, NY 14853, USA}

\begin{abstract}
We present time series photometry of the M~dwarf transiting exoplanet system GJ~436 obtained with the EPOCh (Extrasolar Planet Observation and Characterization) component of the NASA {\it EPOXI} mission. We conduct a search of the high-precision time series for additional planets around GJ~436, which could be revealed either directly through their photometric transits, or indirectly through the variations these second planets induce on the transits of the previously known planet. In the case of GJ~436, the presence of a second planet is perhaps indicated by the residual orbital eccentricity of the known hot Neptune companion. We find no candidate transits with significance higher than our detection limit. From Monte Carlo tests of the time series, we rule out transiting planets larger than 1.5~$R_{\oplus}$ interior to GJ~436b with 95\% confidence, and larger than 1.25~$R_{\oplus}$ with 80\% confidence. Assuming coplanarity of additional planets with the orbit of GJ~436b, we cannot expect that putative planets with orbital periods longer than about 3.4 days will transit. However, if such a planet were to transit, we rule out planets larger than 2.0~$R_{\oplus}$ with orbital periods less than 8.5 days with 95\% confidence. We also place dynamical constraints on additional bodies in the GJ~436 system, independent of radial velocity measurements. Our analysis should serve as a useful guide for similar analyses of transiting exoplanets for which radial velocity measurements are not available, such as those discovered by the {\slshape Kepler} mission. From the lack of observed secular perturbations, we set upper limits on the mass of a second planet as small as 10 $M_{\oplus}$ in coplanar orbits and 1 $M_{\oplus}$ in non-coplanar orbits close to GJ 436b.  We present refined estimates of the system parameters for GJ~436. We find $P$ = $2.64389579 \pm 0.00000080$~$d$, $R_{\star}$ = $0.437\pm 0.016$~$R_{\odot}$, and $R_{p}$ = $3.880 \pm 0.147$~$R_{\oplus}$. We also report a sinusoidal modulation in the GJ~436 light curve that we attribute to star spots. This signal is best fit by a period of 9.01 days, although the duration of the EPOCh observations may not have been long enough to resolve the full rotation period of the star.  
\end{abstract}


\section{Introduction}
{\it EPOXI} (EPOCh + DIXI) is a NASA Discovery Program Mission of Opportunity using the Deep Impact flyby spacecraft \citep{Blume05}. From January through August 2008, the EPOCh (Extrasolar Planet Observation and Characterization) Science Investigation used the HRI camera \citep{Hampton05} with a broad visible bandpass to gather precise, rapid cadence photometric time series of known transiting exoplanet systems. The majority of these targets were each observed nearly continuously for several weeks at a time. In Table 1 we give basic information about the seven EPOCh targets and the number of transits of each that EPOCh observed. 

 One of the EPOCh science goals is a search for additional planets in these systems. Such planets would be revealed either through the variations they induce on the transits of the known exoplanet, or directly through the transit of the second planet itself. This search is especially interesting in the case of the GJ~436 system, since the non-zero eccentricity of the known Neptune-mass planet, first measured by \cite{Butler04}, may indicate the presence of a second planetary companion \citep{Maness07}. Because GJ~436 is a nearby M~dwarf, it is also the only EPOCh target for which we are sensitive to planets as small as 1.0 $R_{\oplus}$. We will describe the searches for additional planets conducted on the remaining EPOCh targets in subsequent papers.

The search for transiting Earth-sized planets in the GJ~436 light curve is scientifically compelling for the following four reasons. First, theoretical predictions of the mass-radius relation for ``super Earths'' are still largely observationally unconstrained, with the exciting exceptions of the two known transiting super Earths CoRoT-7b \citep{Leger09} and GJ~1214b \citep{Charbonneau09}. Depending on the level of observational uncertainty, a measurement of the mass and radius of a super Earth could point to the presence of a large amount of water or iron (enabled with 10\% uncertainty), or allow us to distinguish between a planet composed predominately of water ice, silicates, or iron (enabled with 5\% uncertainty; \citealt{Seager07}). Second, the discovery of two transiting bodies in the same system would permit the direct observation of their mutual dynamical interactions. This would enable constaints on the masses of the two bodies independent of any radial velocity measurement \citep{Holman05, Agol05}. Since radial velocities can only be observed for planets above a certain mass limit, this is an important tool for future surveys of stars too faint for radial velocity measurements. Third, the discovery of an Earth-sized planet at an orbital radius outside that of a giant planet would inform theories of planet formation. Hot Earths are predicted to be captured in low order mean motion resonances with migrating giant planets \citep{Raymond06, Mandell07}. Since the phenomenon of Earth-sized planets at larger orbital radii than Jovian planets is not observed in our own solar system, observations of exoplanet systems are particularly important for this question.

Finally, the eccentricity of the known transiting Neptune-mass planet, GJ~436b \citep{Butler04}, may indicate the presence of an additional perturbing planet, since the assumed circularization timescale for the known planet is much less than the age of the system \citep{Maness07, Deming07, Demory07}. \cite{Ribas08} claimed evidence for a 5 $M_{\oplus}$ super Earth in radial velocity observations of GJ~436, but this proposed planet was ruled out by subsequent investigations \citep{Alonso08,Bean08b}. The absence of this additional perturbing body in the GJ~436 system would also be very scientifically interesting. If no other body is present to explain the eccentricity of GJ~436b, the observed eccentricity requires a very high tidal dissipation parameter, $Q$. The current estimate of the circularization timescale assumes a $Q$ value for the hot Neptune similar to the value derived for the ice giant planets in our own solar system, so a substantially different $Q$ would indicate physical properties of GJ~436b very different from these ice giants \citep{Deming07}. \cite{Jackson08} show that a ratio of planetary tidal dissipation parameter to  planetary Love number $Q/k_{2}$ for GJ~436b greater than $10^{6.3}$ can explain the system's eccentricity (the Love number $k_{2}$ is theoretically between 3/2, in the case of a homogeneous body, and 0, in the case of a centrally condensed body, but ranges between 0.3 and 0.6 for gas giants in the Solar System; \citealt{Bursa92}). In contrast, Uranus and Neptune, the solar system bodies presumably most similar in composition and mass to GJ~436b, have tidal $Q$ parameters estimated at $Q_{U}<3.9\times10^{4}$ and $1.2\times10^{4}< Q_{N}<3.3\times10^{5}$ respectively \citep{Tittemore89, Banfield92}--- several orders of magnitude smaller than the $Q$ necessary to explain the eccentricity of GJ~436b . If the eccentricity is not attributable to a high $Q$, there may instead be an additional perturbing body maintaining the system's eccentricity. The possibility of a close-in resonant companion in 2:1 or 3:1 resonance with GJ~436b is strongly disfavored by transit timing measurements \citep{Pont09}. \cite{Batygin09} proposed possible secular perturbers to GJ~436b, the presence of which would be consistent with observed radial velocities, transit timing measurements, and the non-zero eccentricity of the system. \cite{Bean08b} also quantified the improvement to the goodness-of-fit of the GJ~436 radial velocity data with the addition of perturbing planets to the model---the parameter space they investigated included putative planets of lower mass and eccentricity than those suggested by \cite{Batygin09}. The existence and possible orbital parameters of this putative planet are still open questions. In addition, the recent discovery of the second transiting hot Neptune, HAT-P-11b, also makes this question timely, since the planetary orbit is also eccentric \citep{Bakos10}.

The remainder of this paper is organized as follows. In Section 2, we describe the photometry pipeline we created to produce the time series. In Section 3, we detail the refinement of system parameters and the search we conduct for additional planets around GJ~436, both for additional transits and for dynamical perturbations to GJ~436b. We present a Monte Carlo analysis of the EPOCh observations of GJ~436 and demonstrate the sensitivity to detect a transiting planet as small as 2.0 times the size of Earth for all periods less than 8.5 days with high confidence. We discuss the upper limits on the mass of additional coplanar and non-coplanar planets with periods between 0.5 and 9 days from dynamical constraints. We also discuss the constraints we place on the rotation period of GJ~436. In Section 4, we present our best candidate transit signal, and from the search for additional transits we place upper limits on the radius and mass of the putative planet GJ~436c.

\section{Observations and Data Reduction}
 
We acquired observations of GJ~436 nearly continuously during 2008 May 5 -- 29, interrupted for several hours at approximately 2-day intervals for data downloads. The basic characteristics of the targets and observations are given in Tables 1 and 2. Observations of this type were not contemplated during development of the original Deep Impact mission; the spacecraft was not designed to maintain very precise pointing over the timescale of a transit (Table \ref{tbl-2}). 
Furthermore, the available onboard memory precludes storing the requisite number of full-frame images (1024$\times$1024 pixels).  Hence the observing strategy used 256$\times$256 sub-array mode for those times spanning the transit, and 128$\times$128 otherwise.  This strategy assured complete coverage at transit, with minimal losses due to pointing jitter exceeding the 128$\times$128 sub-array at other times.

\begin{deluxetable}{cccc}
\tabletypesize{\scriptsize}
\tablecaption{EPOCh Targets
\label{tbl-1}}
\tablewidth{0pt}
\tablehead{
\colhead{Name} & \colhead{$V$ Magnitude} & 
\colhead{Number of Transits Observed\tablenotemark{a}} & \colhead{Dates Observed [2008]}
}
\startdata
HAT-P-4 & 11.22 & 10 & Jan 22--Feb 12, Jun 29--Jul 7 \\ 
TrES-3 & 11.18 & 7 & Mar 8--March 10, March 12--Mar 18\\
XO-2 & 12.40 & 3  & Mar 11, Mar 23--Mar 28\\ 
GJ 436 & 10.67 & 8 & May 5--May 29\\
TrES-2 & 11.41 & 9 & Jun 27--Jun 28, Jul 9--Jul 18, Jul 21--Aug 1\\
WASP-3 & 10.64 & 8 & Jul 18--Jul 19, Aug 1--Aug 9, Aug 11--Aug 17 \\
HAT-P-7 & 10.50 & 8 & Aug 9--Aug 10, Aug 18--Aug 31\\
\enddata
\tablenotetext{a}{Some transits are partial.}
\end{deluxetable}

\begin{deluxetable}{cc}
\tabletypesize{\scriptsize}
\tablecaption{Characteristics of the EPOCh Observations 
\label{tbl-2}}
\tablewidth{0pt}
\tablehead{
\colhead{Instrument Parameter} & \colhead{Value}  
}
\startdata
Telescope aperture & 30 cm \\ 
Spacecraft memory & 300 Mb \\
Bandpass & 350-1000 nm \\
Integration time & 50 seconds \\
Pointing jitter  & $\pm$ 20 arc-sec per hour \\ 
Defocus & 4 arc-sec FWHM \\
Pixel scale & 0.4 arc-sec per pixel \\ 
Subarray size & 256$\times$256 pixels spanning transit, 128$\times$128 otherwise \\
\enddata
\end{deluxetable}

We use the existing Deep Impact data reduction pipeline to perform bias and dark subtractions, as well as preliminary flat fielding \citep{Klaasen05}.  The image motion from pointing jitter produces a significant challenge for photometry at our desired level of precision.  The flat field calibration that was obtained on the ground before launch is not successful at the level of precision needed here, because spatial variation of the sensitivity of the CCD has changed modestly since launch in 2005.  Our observing sequences included observations of a green stimulator LED (``stim lamp'') that could be switched on to illuminate the CCD.  The stim lamp illumination is neither flat nor stable in an absolute sense, but its spatial pattern was designed to be stable.  Hence it is useful to define and monitor changes in the flat field response pattern of the CCD.  But since the stim lamp has a different effective wavelength than the stars, it is not a perfect calibrator for flat field changes.  For this reason we also use a 2D spatial-spline fit to the actual data, as a bootstrap flat field method as described below. 

We extract the photometric time series as follows. We determine the position of the star on the CCD using PSF fitting, by maximizing the goodness-of-fit (with the $\chi^{2}$ statistic as an estimator) between an image and a model PSF with variable position, additive sky background, and multiplicative brightness scale factor. We take advantage of a defocused aperture; given our limited on-board memory, defocusing enables us to spread the starlight over more pixels and extend our duty cycle. The PSF itself has a donut-like shape with a roughly 10 pixel FWHM. A model of the PSF is produced from the drizzle of more than 1200 60$\times$60 pixel cutouts, filtered to eliminate cosmic ray hits before drizzle. The final PSF model is sampled to a tenth of a pixel. A bilinear interpolation of this PSF increases the sampling to a hundredth of a pixel, which is the accuracy to which we estimate the position from the $\chi^{2}$ grid. At this point, we perform cosmic ray filtering by removing images from the sample with a larger than 15$\sigma$ outlier in the residuals between image and best-fit PSF model. Because of the high cadence of the EPOCh observations, we simply reject the approximately 30 images ($\approx$0.1\% of the total) containing a cosmic ray overlying the stellar PSF from the time series. 

From each image we subtract a bias determined from the sigma-clipped median of the overclock pixels in each of the four quadrants of the CCD. These are not true pixels that lie on an unilluminated part of the CCD, but rather the bias values read out after the true pixels and then recorded to pixels on the outside of the FITS images. We subtract a bias independently for each quadrant, since the original Deep Impact reduction pipeline does not account for time-dependent bias variations of the CCD. 

We then process the images to remove several sources of systematic error.

1. We scale down the two central rows by a constant value. Due to the CCD read-out electronics, there is a reduction in signal in the pixels at the internal boundary of the two upper and lower imaging regions. However, because the Deep Impact pipeline flat fields these pixels in the same way as the others, we observe the brightness of the star to increase by about 3\% when the stellar PSF lies on the central rows if we do not apply this correction.

2. We scale down the central columns by a separate constant value. We observe an increase in brightness on the order of 0.25\% when the stellar PSF overlies the central columns if this correction is not applied. We interpret the physical origin of this sensitivity variation to be the serial read-out register, which is split in the middle of the CCD to allow the rows to be read out at both ends of the register. 

3. We scale the entire image by a multiplicative factor determined by the size of the sub-array. We gathered observations using two sub-arrays of the 1024$\times$1024 CCD as mentioned above: one in 128$\times$128 mode and another in 256$\times$256. We observed an offset in the average out-of-transit brightness between the two sub-arrays of $8.5\times10^{4}$. We correct for the offset in 256$\times$256 mode by performing the photometric extraction of the time series, determining the decrement in brightness observed in 256$\times$256 mode, uniformly dividing the 256$\times$256 images by this value, and repeating this process until the out-of-transit brightness shows no offset between the two observing modes. 

4. We divide the images by a ``stim'', described above. We first bias-correct the stims using the same prescription as we apply to the images. We then process the stims to remove the asymmetrical illumination pattern, which we model as a plane surface in x and y position.

We then perform aperture photometry on the corrected images. We choose an optimal aperture radius based on analysis of the standard deviation of the out-of-transit time series. We find that this standard deviation is minimized for an aperture radius of 10 pixels, corresponding to twice the HWHM of the PSF. After performing the bias subtraction, scaling of the two middle columns and rows, scaling of the 256$\times$256 sub-array images, and division by a stim, the time series still suffers from significant red noise due to the interpixel sensitivity variations on the CCD. At this point, we implement a 2D spline fit to the data with position by fitting a surface, with the same resolution as the CCD, to the brightness variations on the array. We randomly draw a subset of several thousand out-of-transit and out-of-eclipse points from the light curve (from a data set of $\sim$29,988 points) and find a robust mean of the brightness of the 30 nearest neighbors for each. We then fit a spline surface to these samples, and correct each data point individually by linearly interpolating on this best-fit surface. We use only a small fraction of the observations to create the spline surface in order to minimize the potential transit signal suppression introduced by flat fielding the data by itself. 

To produce the final time series, we iterate the above steps, fitting for the row and column multiplicative factors, the 256 mode scaling factor, and the 2D spline surface that minimize the out-of-transit white noise of the photometric time series. We include one additional step to create the final 2D spline, which is to iteratively remove an overall modulation from the GJ~436 light curve, which has a roughly sinusoidal shape with an amplitude of a few parts in 10$^{4}$. We attribute this modulation to star spots, and discuss this signal in Section 3. After performing the 2D spline, we fit a polynomial to the corrected and binned flux, divide this polynomial from the pre-splined time series, and repeat the spline fit. Otherwise, we expect the modulation signal to introduce red noise to the time series, since we correct for interpixel variations with the assumption that the star's intrinsic brightness is constant outside of times of eclipse and transit.

\begin{figure}[h!]
\begin{center}
 \includegraphics[width=6in]{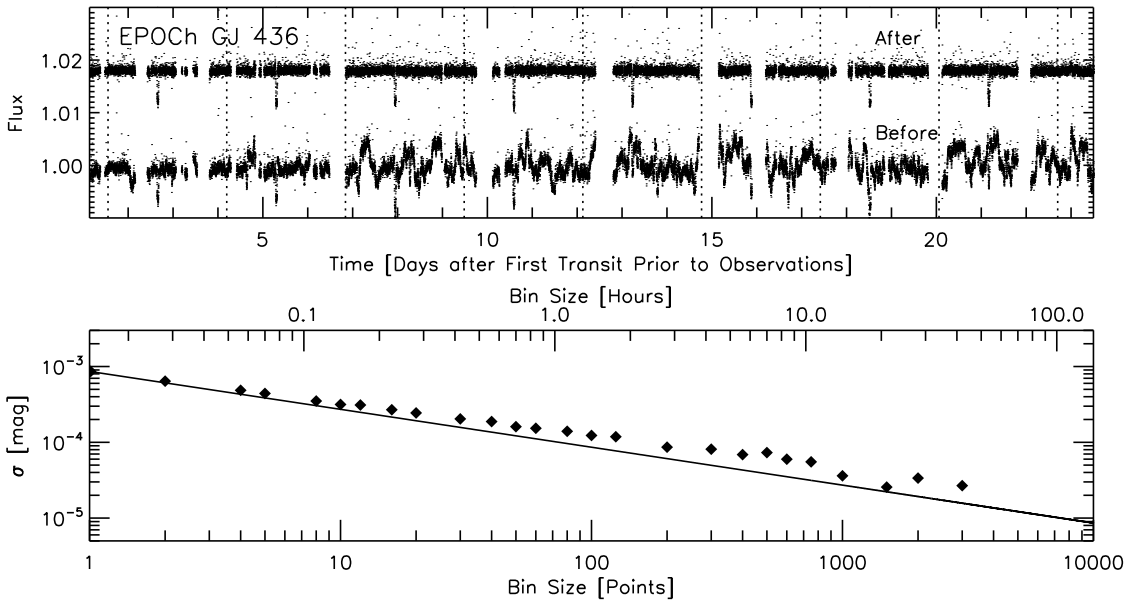} 
 \caption{\textit{Top panel:} GJ~436 time series before (lower curve) and after (upper curve) 2D spline correction. The uncorrected time series (lower curve) has had the two middle rows and columns in each image, and the entire image if taken in 256$\times$256 observing mode, scaled by a multiplicative factor to reduce the flux dependence on position and sub-array size. The images were also divided by a flat field constructed from a stim. We have used the 2D spline to correct for additional interpixel variation in the upper curve. \textit{Bottom panel:} The data (diamond symbols) bin down consistently with the expectation for Gaussian noise (shown with a line, normalized to match the value at N=1).}
   \label{fig:lightcurve}
\end{center}
\end{figure}

After we take these steps to address the systematics associated with the observations, the red noise is largely removed. Figure \ref{fig:lightcurve} shows the GJ~436 time series before and after this 2D spline correction. In the bottom panel, we show that the time series after the 2D spline bins down roughly as predicted for Gaussian noise over timescales less than 4 hours. In the corrected time series, we attribute the scatter to photon noise and low-level cosmic rays. We compare the sigma-clipped standard deviation of the out-of-transit and out-of-eclipse flux to the expected value of the photon noise-limited precision, and find that we are 56\% above the Poisson limit.

We also investigate the transit signal suppression introduced by using a flat field created from the out-of-transit and out-of-eclipse data itself. We avoid the suppression of transits of GJ~436b by excluding those observations from the points used to generate the flat field surface, so that we only use the presumably constant out-of-transit and out-of-eclipse observations to sample the CCD sensitivity. However, if the transit of an additional planet occurs while the stellar PSF is lying on a part of the CCD that is never visited afterward, the 2D spline algorithm instead treats the transit as a dark pixel. To quantify the suppression of additional transits that result from using the 2D spline, we inject transit light curves with periods ranging from 0.5 days to 20 days in intervals of 30 minutes in phase (ranging from a phase of zero to a phase equal to the period) into the EPOCh light curve just prior to the 2D spline step. After performing the 2D spline, we then phase the data at the known injected period and fit for the best radius, using $\chi^{2}$ as the goodness-of-fit statistic. We find that the best-fit radius is suppressed to a mean value of 73\% at all periods, with the standard deviation from that value increasing with period from 3\% at at period of 1 day to 16\% at a period of 10 days. We describe our incorporation of signal suppression into our search for additional planets in greater detail in Section 3.2.

\section{Analysis}

\subsection{Transit Parameters}

We describe here our refinement of the GJ~436 system parameters. When conducting our analysis, we were careful to account for the effects of remaining correlated noise on the parameters and their uncertainties. In the final calibrated GJ~436 light curve, we still observe evidence of correlated noise and trends that have not been corrected by the reduction process. For each transit, we fit a line to the out-of-transit data on both sides of the transit (from 3 minutes outside of transit to half an hour outside of transit) and divide the time series by this line.

We investigate the effects of limb-darkening using several different methods. In the first instance, we use stellar atmosphere models to fix the limb-darkening coefficients to a set of theoretical values. Initially, we use a model atmosphere produced by R. Kurucz \citep{Kurucz94, Kurucz05} corresponding to $T_{\rm eff} = 3500K$, log~$g = 4.5$, [M/H] = 0.0 and $v_{\rm turb} = 0.0$ km/s. We fit the four coefficients of the non-linear limb-darkening law of \cite{Claret00} to 17 positions across the stellar limb. We repeat this fit in 0.2 nm intervals across the {\it EPOXI} bandpass, weighted for the total sensitivity (including filter, optics and CCD response) and photon count at each wavelength. We calculate the final set of coefficients as the average of the weighted sum across the bandpass, for which we find $c_1 = 0.97$, $c_2 = -0.50$, $c_3 = 0.54$, and $c_4 = -0.17$. In order to understand the effect of the stellar atmosphere model choice on the final derived transit parameters, we also generate a set of coefficients from the PHOENIX model \citep{Hauschildt99} with $T_{\rm eff} = 3300K$, log~$g = 4.5$, [M/H] = 0.0 and $v_{\rm turb} = 0.0$ km/s (private communication), and find $c_1 = 0.97$, $c_2 = 0.35$, $c_3 = -0.21$, and $c_4 = -0.03$. As a final test, we also run a parallel analysis where quadratic limb-darkening coefficients are allowed to vary in the fit described below, to see if the data are sufficiently precise to break the degeneracy of the transit shape with geometric parameters and limb-darkening. This is discussed further below.

For each of the three limb-darkening treatments, we fit the transit using the analytic algorithms of \cite{Mandel02}. Using the Levenberg-Marquardt algorithm with $\chi^2$ as the goodness-of-fit estimator, we fit for three geometric parameters: $R_{p}/R_{\star}$, $R_{\star}/a$ and cos $i$. Initially we fix the period to the \cite{Bean08a} value of $P = 2.643904$ days and allow the times of the center of transit to vary independently for each of the eight transits, for a total of eleven free parameters. In addition, there are two quantities that parametrize quadratic limb darkening. These are set as two orthogonal linear combinations of the coefficients $2\gamma_1 + \gamma_2$ and $2\gamma_2 - \gamma_1$. In all three cases we find results that are internally consistent at the 1$\sigma$ level, demonstrating that the choice of limb-darkening treatment does not significantly alter the derived parameters. We note that we can only constrain one linear combination of the limb-darkening coefficients in the quadratic limb-darkening case, the other being degenerate with the geometric parameters to within the precision of the light curve. Due to this degeneracy, the error bars on the geometric parameters are larger in the case where the limb-darkening coefficients are allowed to vary; the uncertainty on the planetary radius $R_{p}$ is larger by a factor of 2.5 and the uncertainty on the stellar radius $R_{\star}$ is larger by a factor of 2.2.

\begin{figure}[h!]
\begin{center}
\includegraphics[width=6in]{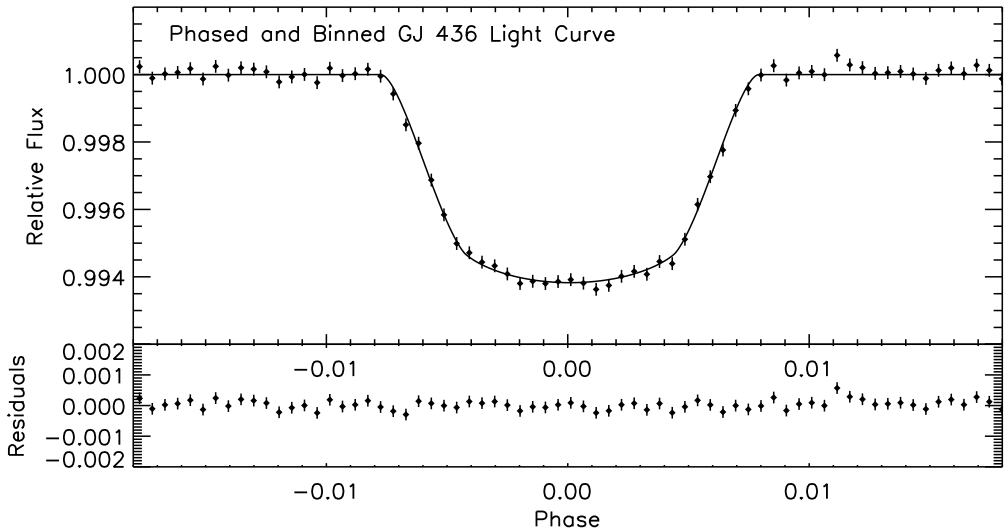}
\caption{Phased and binned GJ~436 light curve with the best-fit model overplotted. The error bars are the rms errors in the bins.}
\label{fig:model}
\end{center}
\end{figure}

For the Kurucz stellar atmosphere limb-darkening coefficients, we find $R_p/R_{\star} = 0.08142 \pm 0.00085$, $R_{\star}/a = 0.0707 \pm 0.0025$ and cos $i = 0.0594 \pm 0.0030$. The light curve is phased and binned into 2 minute intervals with this best-fit model overplotted and shown in Figure \ref{fig:model}. Since there are significant time-correlated systematic errors in the light curve, we calculate the errors on the parameters using the ``rosary bead'' method, in the fashion described by \cite{Winn08}. First, we subtract the best-fit model from the light curve and compile a set of initial residuals. We then shift the residuals along the light curve to the next time stamp in each case, preserving the temporal correlation, add the model back to the residuals, and fit the new light curve as above. For GJ~436, we repeat this process 200 times, which corresponded to a total shift of over three hours, more than three times the duration of the GJ~436 transit. This is sufficient to sample the systematics, which on short timescales vary on the order of 10--30 minutes. The error bars on the parameters are then calculated as the range required to encompass 68\% of the results. Increasing the number of shifts to the residuals does not increase the resulting error bars. We also use the Monte Carlo Markov Chain method, as adapted to transit light curve analysis by e.g. \cite{Holman06}, and find that it results in significantly smaller error bars (30--40\%) than the rosary bead analysis, due to the inability to factor in systematic errors in the light curve. The Monte Carlo Markov Chain method remains a useful tool, however, for assessing the impact of systematic errors on the derived parameters. 

Using the stellar mass from \cite{Torres07}, $M_{\star} = 0.452\pm0.013$, we find $R_{\star} = 0.437\pm 0.016~R_{\odot}$, $R_{p} = 3.880 \pm 0.147~R_{\oplus}$,
 and $i = 86.60 \pm 0.17^{\circ}$. Our measurement of the planetary radius is consistent at the 1$\sigma$ level with the measurements of \cite{Gillon07b}, \cite{Pont09}, and \cite{Southworth08}, and only marginally inconsistent (at the level of 1.3$\sigma$ and 1.4$\sigma$) with the values measured by \cite{Shporer09} and \cite{Gillon07a}. It is significantly smaller than the values obtained by \cite{Torres07}, \cite{Deming07}, and \cite{Bean08a}. Although different treatments of limb darkening may partly account for the discrepancy, in all three cases of limb darkening analysis described above, we find internally consistent results. Our smaller value of the planetary radius would require a reduced mass of the H/He envelope of GJ~436b. \cite{Coughlin08} tentatively suggested that the GJ~436 inclination, transit width and transit depth were increasing with time as the eccentric orbit precessed. Although we do find an slight increase in inclination, again consistent with previous published values, we find a shorter ($58.5\pm2.5$ min) and shallower ($6.173\pm0.037$ mmag) transit than expected from the predicted increase. These findings weaken the trends observed by \cite{Coughlin08} and we therefore cannot confirm any parameter variation with these data.

We fix the times of the centers of transit at the values returned by the least-squares fit described above. These times are shown in Table \ref{tab:times} in UTC. Adding our eight transit times to those published by \cite{Pont09}, \cite{Caceres09}, \cite{Shporer09}, \cite{Gillon07a,Gillon07b}, \cite{Coughlin08}, \cite{Bean08a}, \cite{Deming07} and \cite{Alonso08} we find a new linear, weighted ephemeris of $T_c({\rm BJD}) = 2,454,600.693205 \pm 0.000040$ and $P=2.64389579 \pm 0.00000080$~$d$. 

\begin{deluxetable}{lc}
\tabletypesize{\scriptsize}
\tablecaption{EPOCh Derived Stellar and Planetary Parameters for GJ 436}
\tablewidth{0pt}
\tablehead{\colhead{Parameter} & \colhead{Value}}
\startdata
Light curve Parameters & \\
$P$ (days) & $2.64389579 \pm 0.00000080$ \\
Transit Times (BJD) &  $2,454,592.76134 \pm 0.00022$ \\
& $2,454,595.40458 \pm 0.00026$ \\
& $2,454,598.04822 \pm 0.00021$ \\
& $2,454,600.69271 \pm 0.00045$ \\
& $2,454,603.33657 \pm 0.00017$ \\
& $2,454,605.98079 \pm 0.00035$ \\
& $2,454,608.62326 \pm 0.00026$ \\
& $2,454,611.27015 \pm 0.00040$ \\
$R_p/R_{\star}$ & $0.08142 \pm 0.00085$ \\
$R_{\star}/a $ & $0.0707 \pm 0.0025$ \\
$i$ (deg) & $86.60 \pm 0.17$\\
 & \\
Stellar and Planet Parameters\tablenotemark{a} & \\
$R_{\star}$ ($R_{\odot}$) & $0.437\pm 0.016$\\
$R_{p}$ ($R_{\oplus}$) & $3.880 \pm 0.147$\\

\enddata
\tablenotetext{a}{Using the stellar mass from \cite{Torres07}.}
\label{tab:times}
\end{deluxetable}

\subsection{Search for Additional Transiting Planets}
We perform a robust search of the GJ~436 time series for evidence of additional planets using two methods. The first is a search for additional shallow transits in the light curve. We developed software to search for these additional transits partly modeled on the methods employed on the MOST photometry \citep{Croll07dec,Croll07apr}. The steps involved in the procedure are described in this section.

The photometric precision of the GJ~436 time series (with S/N$\approx$65 for each transit event) should enable a detection of a 1.5 R$_{\oplus}$ planet with S/N$\approx$10 and a 2 R$_{\oplus}$ with S/N$\approx$17, even if the planet produces only a single transit event. In order to test this prediction, we conduct a Monte Carlo analysis to assess how accurately we could recover an injected planetary signal in the GJ~436 light curve. We evaluate our sensitivity to transit signals on a grid in radius and period space sampled at regular intervals in $R_{P}^{2}$ and regular frequency spacing in $P$. We create an optimally spaced grid as follows: for the lowest period at each radius, we determine the radii at which to evaluate the adjacent periods by solving for the radius at which we achieve equivalent signal-to-noise (for this reason, we expect significance contours to roughly coincide with the grid spacing). 
 We use the \cite{Mandel02} routines for generating limb-darkened light curves given these parameters to compute a grid of models corresponding to additional possible planets. If we make the simplifying assumptions of negligible limb darkening of the host star, a circular orbit, and an orbital inclination angle $i$ of 90$^{\circ}$, the set of light curves for additional transiting bodies is a three parameter family. These parameters are radius of the planet $R_{p}$, orbital period of the planet $P$, and orbital phase $\phi$. At each test radius and period, we inject planetary signals with 75 randomly assigned phases into the residuals of GJ~436 EPOCh light curve with the best GJ~436b transit model divided out, and then attempt to recover blindly the injected signal by minimizing the $\chi^{2}$ statistic. We first conduct a coarse $\chi^{2}$ grid search in radius, period, and phase. We select the spacing of this grid to minimize processing time while ensuring that the transit was not missed. We sample the $\chi^{2}$ space at 300 points in period space (at even frequency intervals between 0.5 and 8.5 days), 50 points in radius space (between 0.5 and 5.5 Earth radii) and a variable number of points in phase space from around 30 to nearly 200 (sampled at half a transit duration for each period). We use an expression for the transit duration $\tau$ given by \cite{Seager03}:

\begin{equation}
\mbox{sin }i\mbox{ sin}\left(\frac{\pi\tau}{P}\right)=\sqrt{\left( \frac{R_{\star}+R_{P}}{a}\right)^{2}-\mbox{cos}^{2}i}.
 \label{eq:duration}
\end{equation}

For each test model, we compute the $\chi^{2}$, using the out-of-transit standard deviation to estimate the error in each point. The grid search requires about 24 CPU hours on a 2.66 GHz processor for each radius and period (75 light curves with randomly injected phases). After the grid $\chi^{2}$ minimum is determined, we use the \verb=amoeba= minimization routine \citep{Nelder65} to more finely sample the $\chi^{2}$ space in order to find the $\chi^{2}$ minimum from the specified nearest grid point. We also investigate whether aliases of the best-fit period from the $\chi^{2}$ grid improve the fit. We find that roughly half of the best solutions from the grid are aliases of the injected period, most at either half or twice the value of the injected period, but we test aliases at every integer ratio from 1/35 to 35 times the given period (these other aliases occur less than one percent of the time). We also repeat the finer grid search at the three next lowest $\chi^{2}$ minima, in case the best solution (or an alias of the best solution) lies closer to that grid point. For all 3600 injected signals, we recover a solution which is a better fit (in the $\chi^{2}$ sense) than the exact injected signal. For injected radii greater than 1 $R_{\oplus}$, we recover the period to within 1\% for all injected periods less than 7 days (with the exception of one signal\footnote{In this injected signal with a period of 7 days, only one transit event occurred over the baseline of the observations. Although the radius of the planet corresponding to the single transit event was recovered to 1\%, more than a single transit event is required to accurately recover the period itself.}). For these reasons, we are confident that we are sampling the $\chi^{2}$ space sufficiently finely to locate the best solution. 

\begin{figure}[h!]
\begin{center}
 \includegraphics[width=6in]{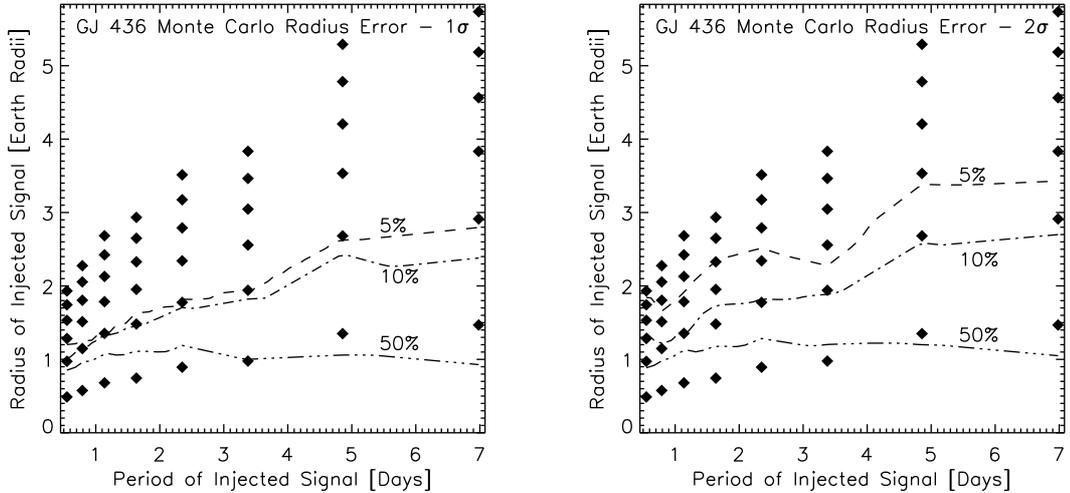} 
 \caption{Constraints on radius from the Monte Carlo analysis. For each point in radius and period, we create 75 light curves with random orbital phases, inject them into the GJ~436 residuals, and attempt to recover them blindly. The diamonds indicate the grid of radii and periods at which we evaluate our sensitivity; the contours are produced by interpolating between these points. At left, contours indicate the relative error in radius (absolute value of recovered-injected/injected radius) that encloses 68\% of the results.  At right, contours indicate the relative error in radius that encloses 95\% of the results. The radius values of the grid points have been upwardly adjusted to incorporate conservative expectations of signal suppression at each period.}
  \label{fig:montecarlo}
\end{center}
\end{figure}

We quantify the success of this analysis by how well the search blindly recovers the known injected transit signal. We define the error on the recovered parameter, for instance period, to be $\mid P_{injected}-P_{observed}\mid/P_{injected}$. We set an approximate error of one sigma for a given parameter at the value that includes 51 of the 75 errors at that point in radius and period (roughly 68\% of all values). Figure \ref{fig:montecarlo} shows this relative error in radius for all searches. We shift the location of the points at which we evaluate our sensitivity, shown as diamonds, to conservatively incorporate the expected level of signal suppression from a planetary transit. As we note in the last paragraph of Section 1, we anticipate suppression of additional transit signals from the bootstrap flat field treatment of the {\it EPOXI} data. We evaluate the suppression we expect at all eight periods on the grid shown in Figure \ref{fig:montecarlo}. We inject planetary transits into the pre-flat fielded light curve in intervals of 30 minutes in phase (from a phase of zero to a phase equal to the period), apply the 2D spline flat field, phase to the known period, and fit for the suppressed transit depth. In this way, we obtain a complete understanding of the suppression at each period, and we can evaluate the level of suppression that we expect with a given confidence. In particular, with 95\% confidence, we expect suppression to be no worse than 71\% the injected radius at 0.55 days, 68\% at 0.79 days, 65\% at 1.13 days, 67\% at 1.63 days, 63\% at 2.35 days, 65\% at 3.38 days, 53\% at 4.86 days, and 55\% at 6.98 days. We incorporate this expected suppression by shifting the effective radius values of the grid points at which we evaluate our sensitivity to additional transits. For example, at 1.63 days, all grid points have been shifted upward in radius by a value of 1/0.67, or 1.49. Because we anticipate no more than 67\% radius suppression at this period (with 95\% confidence), an authentic transit signal would pessimistically appear only 67\% its original radius value after we process the data. For this reason, the recovery statistics corresponding to a 1.0 $R_{\oplus}$ transit depth in the final light curve would be accurate for an original transit signal of a 1.49 $R_{\oplus}$ planet once we fold our expectation of signal suppression. For a planet with 2.0 times the Earth's radius, we are able to recover the radius to better than 5\% accuracy at least 68\% of the time for all periods less than 4 days, and better than 10\% at least 95\% of the time in the same period interval.

\begin{figure}[h!]
\begin{center}
 \includegraphics[width=5.5in]{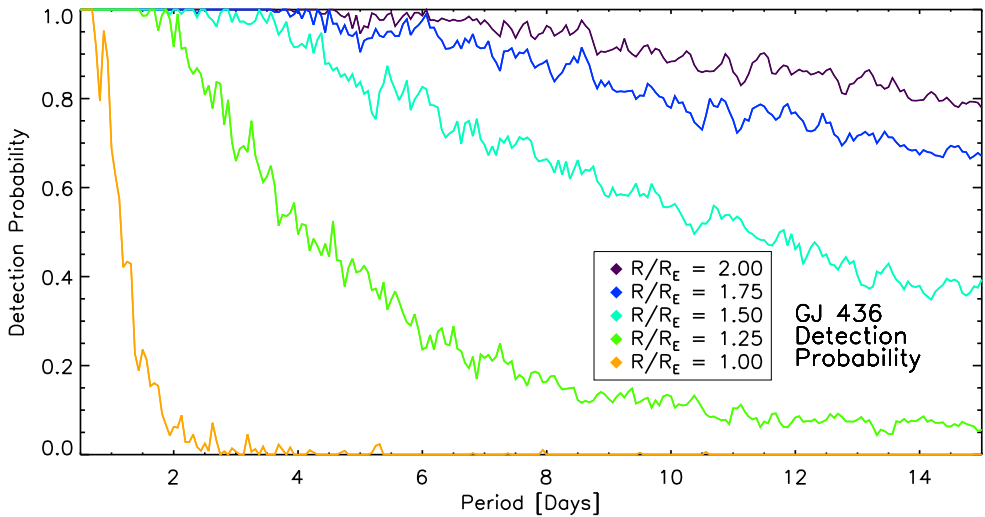} 
 \caption{Detection probability versus period for planets ranging in size from 1 to 2 $R_{\oplus}$. The detection criteria is set by the percentage of phases at a given period for which the number of points observed in transit produces a $\chi^{2}$ improvement of 250, compared to the null hypothesis. We assume a boxcar-shaped transit at the depth of $(R_{P}/R_{\star})^{2}$, where $R_{P}$ at each period and phase includes the expected level of signal suppression.}
   \label{fig:coverage}
\end{center}
\end{figure}

We also evaluate the overall detection probability for putative transiting planets. Given the cadence and coverage of the {\it EPOXI} observations, we determine the number of in-transit points we expect for a given radius, period, and phase (where the phase is evaluated from 0 to 1 periods, in increments of 30 minutes). We then evaluate the expected significance of the detection, assuming a boxcar-shaped transit at the depth of $(R_{P}/R_{\star})^{2}$, and the noise of the actual GJ~436 time series. At each phase and period individually, we scale down $R_{P}$ to incorporate the signal suppression at that ephemeris. We use the improvement in the $\chi^{2}$ over the null hypothesis to define a positive detection, after we have removed the best candidate transit signal (presented in Section 4.1). If we do not first remove this signal, then we are a priori defining a ``detectable'' signal to be any signal more prominent than the best candidate signal, and we would be unable to evaluate this signal's authenticity. We set our detection limit at an improvement in $\chi^{2}$ over the null hypothesis of 250. This level is set by the results of the Monte Carlo analysis; we observe a $\chi^{2}$ improvement of 250 compared to the null hypothesis only at radii and periods at which we correctly recover the injected period (with the exception again of the single signal which presented only one transit event).  Although we use the $\chi^{2}$ as a detection statistic, we do not assign a percent significance to a $\Delta\chi^{2}$ of 250 because of the presence of red noise in the time series. We then determine the fraction of phases for which the signal would be significant enough for {\it EPOXI} to detect, which is shown in Figure \ref{fig:coverage}. For planets larger than about 2.0 times the size of Earth, we would detect the planet with 95\% probability for all periods less than 8.5 days, and we would detect the planet with nearly 100\% probability at periods less than 6 days--- this includes the 2:1 resonance with the known Neptune-mass planet. We would also have a good chance (80\% certainty) of detecting a 1.5 $R_{\oplus}$ planet with a period corresponding to the 2:1 resonance or shorter. Our sensitivity to 1 $R_{\oplus}$ planets is limited to periods of 1.5 days or less for greater than 50\% confidence. Although planets with longer periods are much less likely to transit, we are still sensitive to planets larger than 2.0 $R_{\oplus}$ with 80\% certainty at a period of 15 days, and 70\% certainty at a period of 20 days.


\subsection{Dynamical Constraints on Second Planets}

Over the past few years, many researchers have pointed out that the excellent precision of transit data can reveal second planets.  Such a detection would rely on the deviation of the known planet from a Keplerian orbit; in particular, the orbit and light curve would no longer be perfectly periodic.  We may sort such effects into several categories.  

\begin{figure}[h!]
\begin{center}
\includegraphics[width=6in]{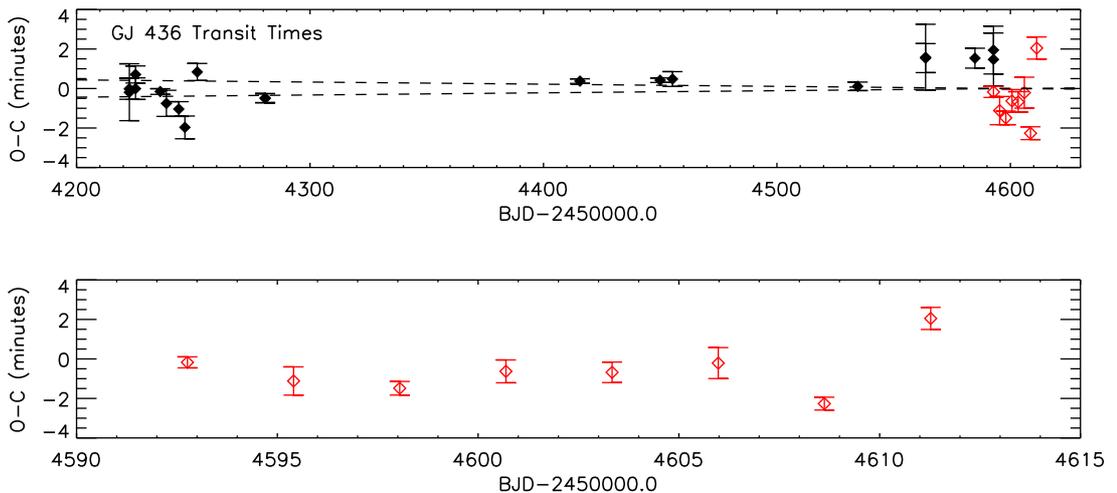}
\caption{Transit timing residuals from the new ephemeris. The solid diamonds are the previously published times from sources listed in the text; the hollow diamonds are the EPOCh times. The propagated error in the new period is overplotted as dashed lines.}
\label{fig:ttvs}
\end{center}
\end{figure}

\begin{figure}
\epsscale{0.75}
\plotone{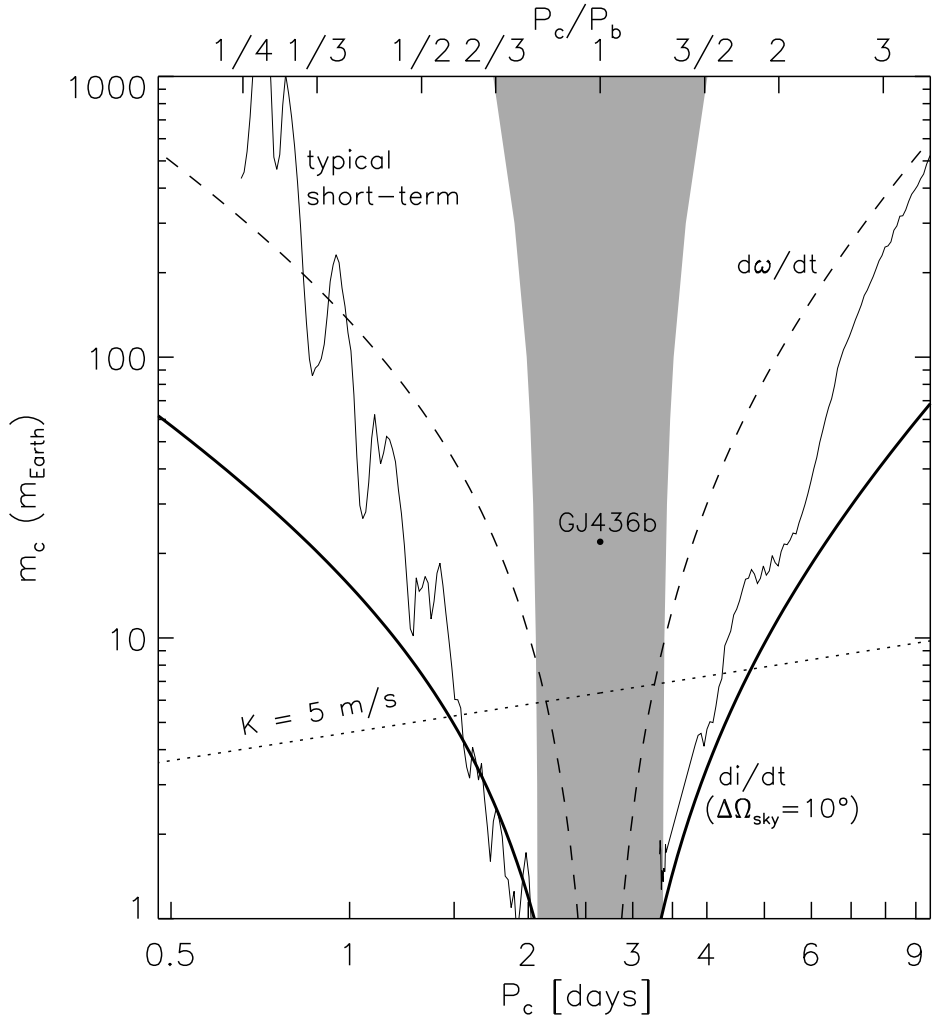}
\caption{ Limits in the mass - period plane for a hypothetical companion planet GJ~436c.  The secular limits are given due to a lack of observed transit inclination and duration change, assuming the companion is currently on a circular orbit.  A coplanar second planet must have a mass below the long-dashed line, due to our upper limit on apsidal motion of b.  A non-coplanar second planet must have $M_2$ below the thick solid line for $|\Delta \Omega_{\rm sky}| \geq 10^\circ$, with the limit scaled up or down in proportion with the actual value of $|\Delta \Omega_{\rm sky}|$.  Such planets could be stable; according to Hill's criterion \citep{1993G} second planets on initially circular orbits are certainly stable outside of the gray area.  The approximate radial velocity limit is given as a critical semi-amplitude $K$ (short-dashed line) that would have been detectable according to the simulations of \cite{Bean08b}.  For perturbers producing short-term fluctuations that would have been detectable in {\it EPOXI} alone, a typical value of the detectable mass is given as a thin solid line; in contrast to the other lines, this should not be viewed as an upper limit (see text).
\vspace{0.1 in}}
\label{fig:seclimits}
\end{figure}

First, on orbital timescales the two (or more) planets exchange small amounts of angular momentum and energy as they approach and recede from mutual conjunctions.  This effect is imprinted on fluctuations of the transit-to-transit orbital period \citep{Agol05, Holman05}, which might be seen in a short string of consecutive transits, like those we are reporting. Each of the 8 EPOCh transit times of GJ~436b is plotted in Figure \ref{fig:ttvs}, and given in Table \ref{tab:times}. We consider variation around a constant period of $>100$~s to be ruled out by our data.  We introduce a small second planet in a series of numerical integrations spanning the transits that {\it EPOXI} measured, to see what mass planet would have been detectable.  For each period of a hypothetical planet of mass $0.01~M_{\rm Jup}$, we chose a circular orbit, coplanar with the known planet, and a grid of initial orbital phases.   We recorded the time of minimum separations between the star and planet b, representing the {\it EPOXI} mid-transit times, and found their standard deviation from a linear ephemeris.  We examined the minimum and the median (on the grid of phases) of those standard deviations.  The minimum was generally considerably higher than the limits from other effects, quoted below.  The median, however, was smaller than $100$~s in orbits close to planet~b.  Even where it was not, as the short-term perturbations are linear in the mass of the perturber, we scale the value of the computed perturbations and the mass to $100$~s and a ``typical'' mass that could have been seen in our data.  We plot that typical mass in Figure~\ref{fig:seclimits} as a thin line.  

Second, on timescales of several tens of orbital periods (for the mass of GJ~436b), orbit-orbit resonances between planets can cause the periods to oscillate, the eccentricities to fluctuate, and the apses to precess.    Our data alone does not have the requisite time span to sample all of such oscillations, although resonant effects do cause the short-term constraints of Figure~\ref{fig:seclimits} to appear jagged, as a function of period.  In combination with light curves taken over several seasons, the data gathered by {\it EPOXI} allow very deep constraints on planets in resonance.  We defer this analysis to future work, as sampling of the relevant timescales is not yet complete using available data.  Similar analyses have already been performed on TrES-1b \citep{2005SA}, HD 189733b \citep{2008MR}, HD 209458b \citep{2007AS,2008MR2}, CoRoT-Exo-1b \citep{2009B}, and TrES-3b \citep{2009G}.  The only considerable difference in the case of GJ~436b is that, because of its small mass, its response to resonant perturbations is proportionally bigger, so the current non-detections of perturbations allow surprisingly low-mass planets to be ruled out.  For instance, once the 6~month timescale is well sampled by transit times with precisions of 7~seconds, then the data could detect or rule out a $\gtrsim0.05M_{\rm Earth}$ planet librating with large amplitude in the 2:1 resonance \citep{Pont09}.  

Third, on timescales of thousands of orbital periods, the planets torque each other's orbits into different orientations and eccentricities, exchanging angular momentum, but not energy.  These secular effects therefore do not change the semimajor axes and periods of the planets, but several recent papers have shown that these changes would still manifest themselves in transit data.  For some effects, an eccentric planet is required and a large impact parameter is desirable, so GJ~436b (with an impact parameter of 0.85) provides a suitable laboratory to search for them.  The remainder of this section evaluates these constraints in detail.

Apsidal motion causes the longitude of transit to move either towards the periastron or towards the apastron of the orbit, depending on $\omega$ and the sign of $\dot{\omega}$ (here $\omega$ is the angle between the ascending node on the plane of the sky and the periastron of the orbit).  The precession period is expected to be considerably longer than the observational baseline, $P_{\rm prec} = 2\pi/\dot{\omega} \gg \tau_{\rm obs}$, so we expect to detect at most a linear change in $\omega$.  One consequence of precession is that the observed period of transits generally differs from the true (sidereal) orbital period.  However, because the true period is known only from radial velocity observations, which are several orders of magnitude less precise than transit data for this purpose, in practice this method is not constraining.  On the other hand, the period of transits would differ from the period of occultations, so multi-epoch secondary eclipse observations should be able to place a tight constraint on $\dot{\omega}$ from this technique \citep{2002M,2007HG}.  Currently, the technique with the highest precision is to use transit \emph{durations} to constrain $\dot{\omega}$, as the duration changes linearly in time depending on the true planet-star separation at transit.  This dependence comes in two forms: (1) the impact parameter is directly proportional to this separation and (2) the tangential velocity is inversely proportional to this separation, according to Kepler's second law.  It has been shown \citep{1959K} that these two effects cancel, for linear changes in $\omega$, at an impact parameter of $b=1/\sqrt{2}$.  At the impact parameter of GJ~436b, the dependence of the impact parameter is more important, as has been emphasized by \cite{Shporer09}.  Both effects together cause the duration to change at a rate:
\begin{equation}
\frac{d \omega}{dt} = \frac{dT/dt}{T} \frac{1 + e \sin{\omega} }{ e \cos \omega } \frac{1 - b^2}{ 1-2b^2 }  \label{eqn:omdot}
\end{equation}
 \citep{2008PK, 2008JB}, where $T$ is the duration between the times when the projected centers of the planet and star are separated by $R_\star$.  We combined the transit duration of our combined light curve, considered as a ``single epoch'' measurement of extremely high quality, with the transit durations reported over the last two years.  Here and below, we take the values from the compilation and homogeneous analysis of \cite{Coughlin08}.\footnote{ \cite{Coughlin08} analyzed the data in two ways; we use the results assuming fixed stellar and planetary radii for all the data sets.}  We measure $dT/dt = (1.9 \pm 3.7) \times10^{-3}$~min$/$day, so infer $\dot{\omega} = (-8 \pm 15)\times 10^{-3}$~deg$/$day by equation~\ref{eqn:omdot}.  Taking $|\dot{\omega}| \lesssim 4.5 \times 10^{-2}$~deg$/$day to be a conservative upper limit, we now consider what constraints this puts on additional planets in the GJ~436 system.  As \cite{2007HG} pointed out, the asymptotic formulae \citep{2002M} for apsidal rate due to second planets underestimates their strength by a factor of $> 2$ when the perturbing planet is within a factor of $2$ in semi-major axis of the transiting planet.  Thus we use the full formula for apsidal motion caused by a perturbing planet on a circular orbit \citep{2007HG,1999MD}:
 \begin{equation} \label{eq:domam}
 \frac{d \omega}{dt} = \frac{\pi}{2 P_b} \frac{M_c}{M_\star} \times \left\{
 \begin{array}{cl} 
 (a_c/a_b)   b_{3/2}^{(1)}(a_c/a_b)  &  a_c<a_b \\
(a_b/a_c)^2 b_{3/2}^{(1)}(a_b/a_c)   & a_c>a_b
  \end{array} \right\}
 \end{equation}
where 
 \begin{equation}
b_{3/2}^{(1)}(x)= \frac1\pi \int_0^{2\pi} \frac{ \cos \psi d \psi }{ (1 + 2 x \cos \psi + x )^{3/2} }
\end{equation}
is the Laplace coefficient, which we compute numerically.  We may invert this formula to obtain a constraint on the mass of a second planet, as a function of its semimajor axis, given the previously quoted constraint on $\dot{\omega}$; this is plotted in Figure~\ref{fig:seclimits}.  Perturbing planets on currently eccentric orbits can torque planet~b at somewhat different rates, depending on the relative apsidal orientation, an effect we neglect here.
 
Another secular effect is nodal precession, which changes both the node and the inclination of the orbit.  The former is measurable in principle by comparing Rossiter-McLaughlin (R-M) observations spanning many years, but GJ~436 is a slow rotator ($P_{\rm rot} \approx 48$~days; \citealt{Demory07}) and thus not a favorable R-M target.  The latter, however, is exquisitely measurable due to the large impact parameter of GJ~436b.  In fact, detections of an inclination change have already been suggested \citep{Ribas08, Coughlin08}, which would imply a several Earth-mass planet in a nearby orbit, and the specific orbits and masses of allowable planets have been debated \citep{Alonso08,2009D, Bean08b}.  We measured $di/dt = (-6 \pm 31)\times 10^{-5}$~deg$/$day by producing a weighted linear ephemeris to our inclination and the ones presented in \cite{Coughlin08}, \cite{Gillon07a,Gillon07b}, \cite{Bean08a}, \cite{Deming07}, \cite{Shporer09}, and \cite{Alonso08}. We find a a conservative upper limit of $|di/dt| \lesssim 9 \times 10^{-4}$~deg$/$day.  The shape of the light curve is much more sensitive to inclination change than to apsidal motion. Next, we convert this to a mass constraint on second planets. The secular theory of \citet[\S 7.2]{1999MD}, gives the rate of inclination change due to a perturbing planet, as the nodes of the planets precess:

\begin{equation}
\frac{di}{dt} = - \frac{d \omega}{dt} \Delta\Omega_{\rm sky}, \label{eq:np}
\end{equation}
where $d\omega/dt$ is given by Equation~\ref{eq:domam} and $\Delta\Omega_{\rm sky}$ is the ascending node of the second planet relative to the ascending node of the known planet, measured clockwise on the plane of the sky.  (This latter quantity is the sky-projected mutual inclination, similar to the sky-projected spin-orbit angle from the R-M literature.)  Equation~\ref{eq:np} assumes small eccentricities and mutual inclination.  We have tested it against numerical integrations for the known planet and a second planet that would be marginally detectable and is far from resonance ($M_c = 35 M_\earth$, $P_c=2.818 P_b$), and find it to be accurate to 20\% over the timespan of the data if the perturbing planet begins on a circular orbit and is inclined less than $20^\circ$ from the orbit of planet~b.  The resulting mass limit is plotted in Figure~\ref{fig:seclimits} for $|\Delta \Omega_{\rm sky}| \geq 10^\circ$.  Because of the lightcurve's sensitivity to inclination change, only a small $\Delta\Omega_{\rm sky}$ is needed for an inclination change over a certain angle to be as detectable as apsidal motion through the same angle.  For $\Delta\Omega_{\rm sky} = 0.37^\circ$, the inclination constraint has the same magnitude as the apsidal motion constraint.  According to Equation~\ref{eq:np} it also has the same form.  

We note a degeneracy between the effects of apsidal motion and nodal precession.  The light curve is most sensitive to both via transit duration change, so with a judiciously chosen orbit for a second planet, the two effects mostly cancel at the present time \citep{2002M}.  In such a case, the precession of the planet in its plane causes the impact parameter to increase at the same rate the precession of the planet out of its plane causes the impact parameter to decrease.  This happens for $\Delta\Omega_{\rm sky} = -0.37^\circ$, so a much larger $M_c$ is possible for that particular orbit.  The degeneracy could be broken by the shape of the light curve, as the changing velocity accompanying apsidal motion is not canceled, so the change in the ingress/egress duration and the depth of the transit due to the limb-darkened star may be observable.  However, in practice the radial velocity constraint provides a stronger limit for such orbits.

We have also plotted as a gray area in Figure \ref{fig:seclimits} the orbits near GJ~436b that are not guaranteed to be stable by Hill's criterion (e.g., \citealt{1993G}).  We took account of the eccentricity of b, so the stability criterion is approximately
\begin{equation}
\frac{a_{\rm outer} - a_{\rm inner}}{a_{\rm inner}} > \sqrt{ \frac{8}{3} e_{b}^2 + 5.76 \Big( \frac{M_{\rm inner} + M_{\rm outer}}{M_\star} \Big)^{2/3} }.  
\end{equation}

There is only a narrow region that both satisfies the stability criterion and is ruled out by these secular constraints more strongly than by the radial velocity measurements.  However, these dynamical constraints will certainly tighten in the future with both longer observing baselines and the measurement of the occultation period, to be compared with the transit period.  The secular limits given above will improve approximately linearly in time, until 1 radian of the secular cycle has elapsed.  This timescale may be computed using Equation~\ref{eq:domam} with max$(M_b, M_c)$ substituted for $M_b$, and is typically decades.  In contrast, most relevant short-term and resonant timescales are shorter than the several-year time span of data collected so far, so apart from sampling all the timescales, those constraints have saturated.  Thus, they will only improve considerably with improved precision of transit data.

In the case of GJ~436, excellent radial velocity data was in hand before transits were ever discovered, so it is not surprising that radial velocity constraints are stronger than the transit timing constraints for a wide range of periods of perturbers.  However, soon exquisite transit data will be available from the {\slshape Kepler} mission, with little or no useful constraints from radial velocity.  Therefore, the techniques illustrated here for GJ~436 are expected to be very useful in those cases.

Finally, there are still several potentially confounding effects which are well below the sensitivity of the data.  The apsidal line may precess not by a second planet, but by either the post-Newtonian relativistic correction, by a tidal bulge raised on the planet \citep{2009RW}, or by rotational oblateness of the star \citep{2002M}.  These effects have expected magnitude $6.5\times 10^{-5}$~deg$/$day \citep{2008PK, 2008JB}, $1.0\times 10^{-6}$~deg$/$day \citep{2009RW}, and $3\times 10^{-8}$~deg$/$day  \citep{2002M} respectively, all far below the current reach of the data.  Also, nodal precession may occur and be observable if the stellar spin axis and the orbit normal do not lie in the same angle as projected on the plane of the sky.  Such precession is driven by rotational oblateness, so its timescale is similar to apsidal motion by the same mechanism, approximately $3\times 10^{-8}$~deg$/$day.

 We note also that a detailed search for Trojan asteroids to the known planet GJ~436b using the transit timing method proposed by \cite{Ford06} could place interesting limits on Trojans in this system, although we do not specifically address this question in this work. However, Trojan bodies could also produce a photometric transit, so our constraints on additional transiting bodies could be applied to the detection of Trojans as well. These Trojan asteroids would have a period equal to GJ~436b (trailing or preceding the hot Neptune planet at the fixed phase set by the L4 and L5 Lagrange points), so without conducting a detailed analysis specific to Trojan asteroids, we conservatively would have been able to detect Trojans that produced a transit depth equivalent to a 1.5 $R_{\oplus}$ planet with nearly 100\% certainty from the detection probabilities shown in Figure \ref{fig:coverage}.

\subsection{Rotation Period of GJ~436}
\begin{figure}[h!]
\begin{center}
 \includegraphics[width=6in]{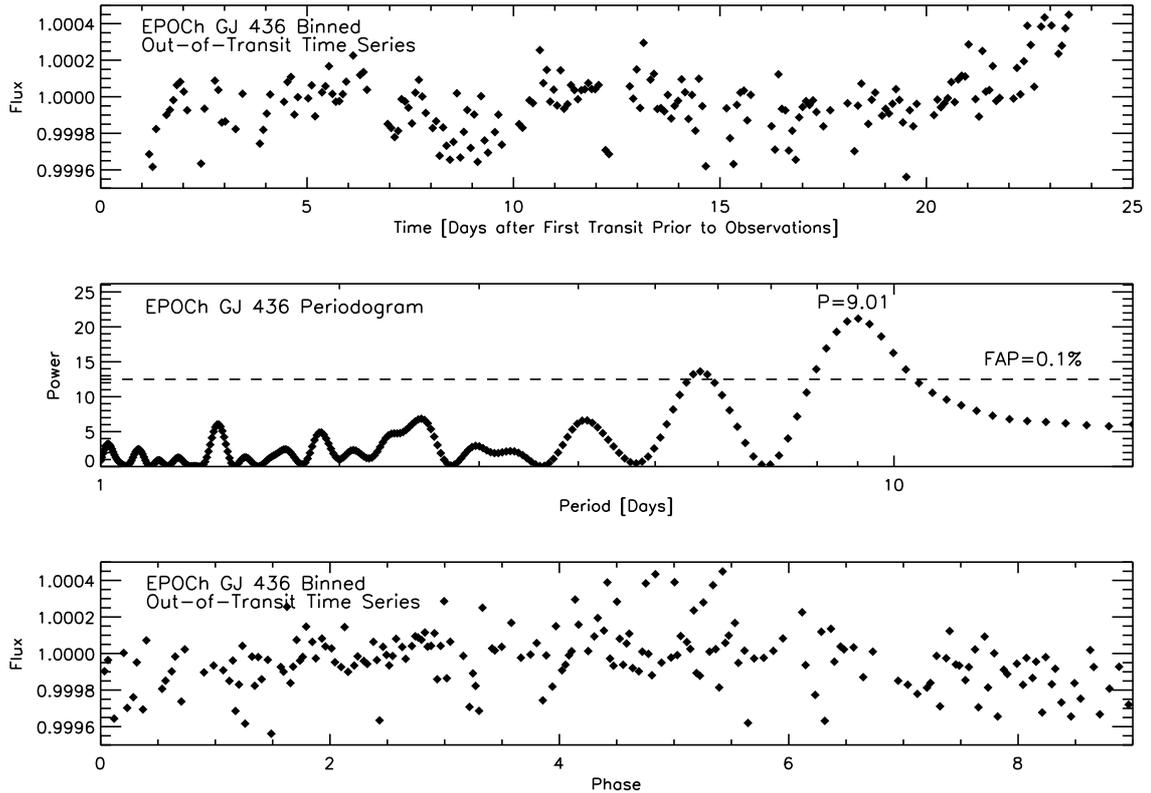} 
 \caption{\textit{Top panel:} GJ~436 out-of-transit and out-of-eclipse EPOCh observations, binned in two hour increments. \textit{Middle panel: } Lomb-Scargle periodogram of binned observations, with false alarm probability of 0.1\% overplotted. The most significant peak corresponds to a period of 9.01 days. \textit{Bottom panel: } Time series phased to most significant period of 9.01 days.}
   \label{fig:rotate}
\end{center}
\end{figure}

As previously stated, we fit a sinusoidal modulation in the GJ~436 light curve with a polynomial and remove it in order to conduct the search for additional transits. Here we consider the astrophysical significance of this signal, which we attribute to changes in the apparent brightness of the star due to star spots. The sinusoidal modulation has a peak amplitude near 0.5 millimagnitudes and is shown in the top panel of Figure \ref{fig:rotate}.  \cite{Butler04} give an upper limit on $V$~sin~$i$ of 3~km~s$^{-1}$; if we assume rigid body rotation and a stellar radius from this work of $R_{\star} = 0.437\pm 0.016~R_{\odot}$, we infer a lower limit on the rotation period of $\approx$ 7.4 days. We investigate the periodicity of the EPOCh GJ~436 observations by binning the time series in two hour increments and creating a Lomb-Scargle periodogram of the binned observations. We find only one peak with false alarm probability significantly less than $0.01\%$ at $P_{\rm rot}$=9.01 days;  the time series is shown phased to this period in the bottom panel of \ref{fig:rotate}. The best-fit rotational period of 9.01 days is longer than the lower limit set by $V$~sin~$i$. However, \cite{Demory07} observed flux variations of GJ~436 nearly 20 times larger with amplitude of 1$\%$ which were best fit by a stellar rotation period $\approx$ 48 days. It is therefore possible that the baseline of the {\it EPOXI} observations is not long enough to resolve the full rotation period of the star, and the modulation we observe is due to multiple starspots passing into view over a period shorter than the rotation period of the star.

\section{Discussion}

\subsection{Best Candidate Transit Signal}

\begin{figure}[h!]
\begin{center}
 \includegraphics[width=6in]{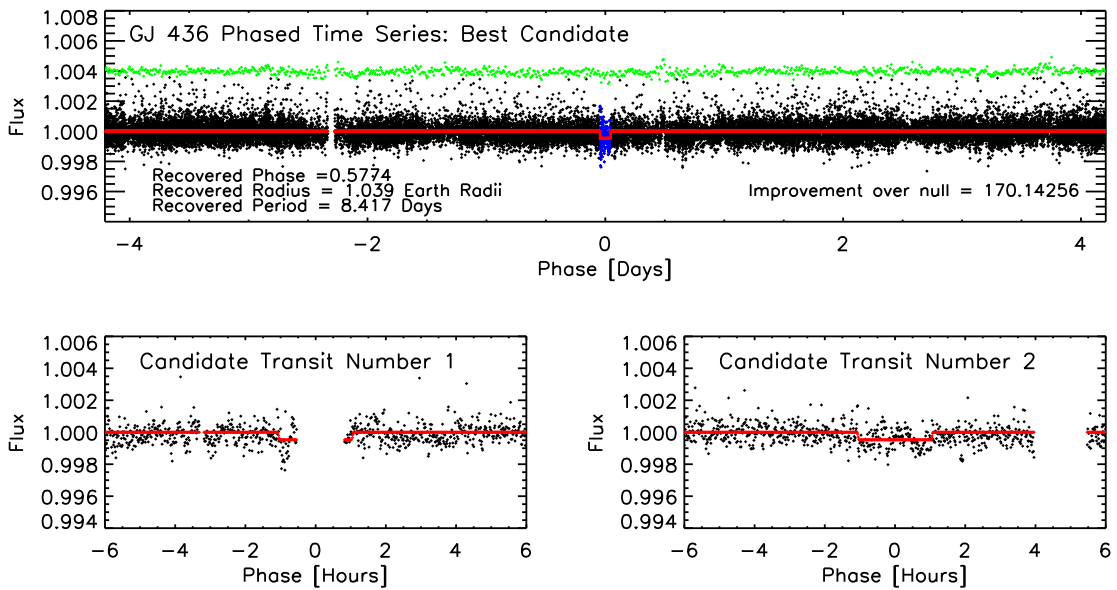} 
 \caption{\textit{Top panel:} Best transit candidate, shown phased (black points) with best-fit model overplotted in red. The green curve directly above is the phased light curve binned by a factor of 30.
 \textit{Bottom panels:} The two observed candidate transit events which contribute to the best candidate signal, with best-fit model overplotted in red. The transit event at left contributes an improvement over the null $\chi^{2}$ of 49, and the right transit contributes an improvement in $\chi^{2}$ of 121. Since we place our significance limit at $\Delta\chi^{2}=250$, we cannot claim a positive detection.}
   \label{fig:best_candidate}
\end{center}
\end{figure}

We find no transit signals for GJ~436c at the significance limit set by $\Delta\chi^{2}$=250. We present our best candidate here, which corresponds to a planet with radius 1.04 $R_{\oplus}$ and period 8.42 days. The improvement in the $\chi^{2}$ over the null hypothesis is 170. We investigate the possibility of signal suppression by masking these suspected transit points from the points used to create the 2D spline, and recreating the spline surface. We find that the candidate transit depth is unaltered, indicating that these points are well-sampled on the CCD. In Figure \ref{fig:best_candidate}, we show the phased and binned best candidate signal, as well as the two observed candidate transit events separately. The gap in the middle of the first event is due to the star wandering off the bottom edge of the CCD for a period of about an hour. 

We also perform an identical transit search on a flipped version of the time series (we subtract 1 from the normalized time series and multiply by -1), since we expect red noise fluctuations to introduce both positive and negative imposter transit signals. We find that the best solution to the inverted time series produces an improvement over the null hypothesis of $\Delta\chi^{2}=106$.

\subsection{Radius constraints}

From the results of our Monte Carlo analysis and phase coverage analysis, we rule out transiting planets orbiting interior to GJ~436b larger than 1.5 Earth radii with 95\% confidence. We would expect to detect such a planet with 95\% certainty, and recover its radius to within 15\% with 95\% confidence (referencing Figures \ref{fig:coverage} and \ref{fig:montecarlo}, respectively). For planets exterior to the orbit of GJ~436, we can no longer assume that additional planets will transit the host star---since GJ~436b itself has an inclination of $i = 86.60 \pm 0.17^{\circ}$, and the host star has a radius of $R_{\star} = 0.437\pm 0.016~R_{\odot}$ (from this work), we would anticipate coplanar planets to transit only with periods less than about 3.4 days. However, the orbital inclination of Earth differs from the orbital inclinations of the gas giants by up to 2.5$^{\circ}$ in the case of Saturn \citep{Cox00}. If the putative GJ~436c had an inclination which was 2$^{\circ}$ closer to edge-on than GJ~436b, we should be able to detect transits out to nearly 13 days. For all periods less than about 8.5 days, we would have detected the transit of a planet 2.0 $R_{\oplus}$ or larger with 95\% probability if the planet produced a transit (shown in Figure \ref{fig:coverage}). This includes the 2:1 resonance with GJ~436b at about 5.3 days, although NICMOS observations of the transit times of GJ~436b, with timing variations less than a few seconds, disfavor planets in the 2:1 resonance-- a planet in the 2:1 resonance with GJ~436b with a mass as small as $0.01$ $M_{\oplus}$ should produce at least 7 second variations \citep{Pont09}. For periods less than about 20 days, we would have detected the transit of a planet of 2.0 $R_{\oplus}$ or larger with 70\% probability.

\subsection{Mass constraints}

We use the theoretical mass-radius relationships for super Earths and hot Neptunes calculated by \cite{Fortney07} to place approximate mass constraints, given our radius constraints derived from the search for additional transiting planets. We rule out planets larger than 1.5 $R_{\oplus}$ interior to GJ~436b--- using the analytic formulae given in \cite{Fortney07}, such a planet would have a mass around 0.8 $M_{\oplus}$ if it were pure ice, 2.9 $M_{\oplus}$ if it were pure rock, and 4.6 $M_{\oplus}$ if it were pure iron. At semi-major axes larger than that of GJ~436b, since additional planets may not transit, we are unable to set firm upper limits from the lack of transits. If such a planet with a period less than 8.5 days did transit, we would be sensitive with 95\% confidence to radii as small as 2.0 $R_{\oplus}$-- even at a period of 20 days, although the transit probability is much less likely, we would still detect a planet this size with 70\% certainty. We can therefore rule out transiting planets at periods less than 8.5 days with masses greater than 2.3 $M_{\oplus}$ assuming a pure ice composition, 9.6 $M_{\oplus}$ assuming a pure rock composition. Although a pure iron planet with a radius of 2.0 $R_{\oplus}$ would have a mass of 63.5 $M_{\oplus}$ \citep{Fortney07}, this composition is perhaps unrealistic, even assuming a formation history with mantle-stripping collisions \citep{Marcus10}. The maximum mass  of a 2.0 $R_{\oplus}$ planet, using the relations derived by \cite{Marcus10} for maximum collisional stripping, would be closer to 20 $M_{\oplus}$. Planets with periods less than 7 days have been ruled out in the GJ~436 system by radial velocity constraints down to about 8 $M_{\oplus}$ with 3$\sigma$ confidence \citep{Bean08b}. Our limits on the presence of pure rock planets are therefore complementary with previous, stronger constraints. 

We find that the EPOCh dynamical constraints on additional planets with periods from 0.5 to 9 days rule out coplanar secular perturbers as small as 10 $M_{\oplus}$ and non-coplanar secular perturbers as small as 1 $M_{\oplus}$, as shown in Figure \ref{fig:seclimits}. These dynamical constraints are not as strong as current radial velocity constraints, except in orbits very close to that of GJ~436b. However, we anticipate that dynamical analyses similar to those presented in this work will prove useful to the community in cases of planets with masses below current radial velocity detectability, such as those that {\slshape Kepler} will find.

\subsection{Eccentricity of GJ~436b}

 The eccentricity of GJ~436b has been attributed to two possible mechanisms. First, the residual eccentricity can be attributed to excitation from the dynamical interactions of GJ~436b with an as-yet undetected additional planet. \cite{Bean08b} tested this second hypothesis by finding how well the radial velocity data could be improved by adding perturbers to the system and evolving the system forward by numerically integrating the Newtonian equations of motion. Their analysis ruled out perturbers greater than 8 $M_{\oplus}$ at periods less than about 11 days (semi-major axes less than 0.075 AU) with high confidence. They presented radial velocity solutions that improved the fit by up to 4$\sigma$ at smaller masses with periods between 4 and 11 days. We rule out rocky transiting bodies down to 9.6 $M_{\oplus}$ with periods less than 8.5 days with 95\% confidence in the GJ~436 system. However, this doesn't preclude the possibility of an additional planet at these periods which does not transit. \cite{Batygin09} compiled a list of possible dynamically stable secular perturbers which are consistent with the transit times, radial velocities, and observed eccentricity of GJ~436b. These proposed additional planets all have periods greater than 16 days, however. {\it EPOXI} would be sensitive to a transiting planet larger than 2.0 $R_{\oplus}$ with close to 80\% probability at 15 days (see Figure \ref{fig:coverage}).

 The second possible explanation for the eccentricity of GJ~436b is a tidal $Q$ parameter that is much larger than that of the ice giants in our solar system (thereby increasing the circularization time to greater than the age of the system)---such a $Q$ would need to be 1--2 orders of magnitude larger than that measured for Neptune \citep{Batygin09, Banfield92}. However, the tidal $Q$ for Jupiter may be as high as $2\times10^{6}$ with the assumption that Jupiter and Io are orbiting in a steady-state configuration; and may be even higher if that assumption is false (and tides on Io are currently dominant) \citep{Jackson08}. In fact, \cite{Oglivie04} find that hot Jupiters may typically have $Q$ values near $5\times10^{6}$, so a value of $10^{6.3}$ for GJ~436b for $Q/k_{2}$, proposed by \cite{Jackson08}, may not be unreasonable (the Love number $k_{2}$ is typically near 0.5 for Solar System gas giants; \citealt{Bursa92}). However, \cite{Batygin09} suggest that the actual $Q$ for GJ~436b must be higher still.

A definitive explanation for the eccentricity of GJ~436b is so far undetermined, but a resolution to this question is observationally tractable. \cite{Batygin09} provide a thorough discussion of follow-up observations that could measure the signal of a secular perturber to GJ~436b: radial velocity measurements are sensitive enough at their current level to resolve the periodogram signature of such a perturber, and a long baseline of transit times at the level of $\Delta t<$ 10 s could also confirm its presence \citep{Batygin09}. If these methods find no signal corresponding to a perturbing body, then the tidal $Q$ of GJ~436b may indeed be much higher than those for the ice giants in our solar system. The explanation proposed by \cite{Jackson08} may alternatively be correct-- the $Q$ values for solar system giant planets are current underestimations of the true $Q$ value for these planets, in which case the tidal $Q$ necessary to explain GJ~436b may not be inconsistent with that of Neptune.

\section{Acknowledgments} We are extremely grateful to the {\it EPOXI}  Flight and Spacecraft Teams that made these difficult observations possible.  At the Jet Propulsion Laboratory, the Flight Team has included M. Abrahamson, B. Abu-Ata, A.-R. Behrozi, S. Bhaskaran, W. Blume, M. Carmichael, S. Collins, J. Diehl, T. Duxbury, K. Ellers, J. Fleener, K. Fong, A. Hewitt, D. Isla, J. Jai, B. Kennedy, K. Klassen, G. LaBorde, T. Larson, Y. Lee, T. Lungu, N. Mainland, E. Martinez, L. Montanez, P. Morgan, R. Mukai, A. Nakata, J. Neelon, W. Owen, J. Pinner, G. Razo Jr., R. Rieber, K. Rockwell, A. Romero, B. Semenov, R. Sharrow, B. Smith, R. Smith, L. Su, P. Tay, J. Taylor, R. Torres, B. Toyoshima, H. Uffelman, G. Vernon, T. Wahl, V. Wang, S. Waydo, R. Wing, S. Wissler, G. Yang, K. Yetter, and S. Zadourian.  At Ball Aerospace, the Spacecraft Systems Team has included L. Andreozzi, T. Bank, T. Golden, H. Hallowell, M. Huisjen, R. Lapthorne, T. Quigley, T. Ryan, C. Schira, E. Sholes, J. Valdez, and A. Walsh.

Support for this work was provided by the {\it EPOXI} Project of the National Aeronautics and Space Administration's Discovery Program via funding to the Goddard Space Flight Center, and to Harvard University via Co-operative Agreement NNX08AB64A, and to the Smithsonian Astrophysical Observatory via
Co-operative Agreement NNX08AD05A.

DF gratefully acknowledges support from the Michelson Fellowship, supported by the National Aeronautics and Space Administration and administered by the Michelson Science Center.

\end{document}